\documentclass[a4paper]{spie}  

\usepackage[]{graphicx}
\usepackage{amssymb,amsmath,psfrag,url}

\newcommand{\MyMatrix}[1]{\mathrm{#1}}

\newcommand{\MyField}[1]{{\bf{#1}}}

\newcommand{\MyTensor}[1]{{{{#1}}}}
\newcommand{\MySField}[1]{{{#1}}}

\title{FEM investigation of leaky modes in hollow core photonic crystal fibers} 

\author{
Jan Pomplun\supit{\,ab},
Ronald Holzl{\"o}hner\supit{\,c},
Sven Burger\supit{\,ab},  
Lin Zschiedrich\supit{\,ab},
Frank Schmidt\supit{\,ab}, 
\skiplinehalf
\supit{a}
Zuse Institute Berlin,
Takustra{\ss}e 7,
D\,--\,14\,195 Berlin,
Germany
\smallskip\\
\supit{b}
JCMwave GmbH,
Haarer Stra{\ss}e 14a,
D\,--\,85\,640 Putzbrunn, 
Germany
\smallskip\\
\supit{c}
European Southern Observatory,
Karl-Schwarzschild-Stra{\ss}e 2, 
D\,--\,85\,748 Garching,
Germany,\smallskip\\
}

\authorinfo{
Corresponding author: J. Pomplun\\
URL: http://www.zib.de/nano-optics/\\
Email: pomplun@zib.de
}

\begin{document} 
  \maketitle 

\noindent
Copyright 2007  Society of Photo-Optical Instrumentation Engineers.\\
This paper has been published in Proc.~SPIE {\bf 6480}, 6480-22
(2007),  
({\it Photonic Crystal Materials and Devices VI, A. Adibi, S.-Y. Lin, A. Scherer, Eds.})
and is made available 
as an electronic reprint with permission of SPIE. 
One print or electronic copy may be made for personal use only. 
Systematic or multiple reproduction, distribution to multiple 
locations via electronic or other means, duplication of any 
material in this paper for a fee or for commercial purposes, 
or modification of the content of the paper are prohibited.

\begin{abstract}
Hollow-core holey fibers are promising candidates for low-loss guidance of light in various applications,
 e.g., for the use in laser guide star adaptive optics systems in optical astronomy. 
We present an accurate and fast method for the computation of light modes in arbitrarily shaped waveguides. 
Maxwell's equations are discretized using vectorial finite elements (FEM). 
We discuss how we utilize concepts like adaptive grid refinement, higher-order finite elements, and 
transparent boundary conditions for the computation of leaky modes in photonic crystal fibers. Further, 
we investigate the convergence behavior of our methods. 

We employ our FEM solver to design hollow-core photonic crystal fibers (HCPCF) whose cores are formed from
 19 omitted cladding unit cells. We optimize the fiber geometry for minimal attenuation using multidimensional optimization taking into account radiation loss (leaky modes).
\end{abstract}


\section{Introduction}
Photonic crystal fibers (PCF) have a core surrounded by a periodic arrangement of holes and struts.
 This structure prevents leakage of light to the exterior \cite{RUS03}. In contrast to ordinary index-guiding 
optical fibers the core can have a smaller refractive index than the surrounding material or even be hollow.
 Although hard to fabricate there are many application areas where the guidance principles of PCFs offer
 great advantages. An important example for the use of hollow-core photonic crystal fibers (HCPCF)
 is the field of high-power light transmission \cite{RUS03}, e.g. for pulsed lasers for 
adaptive optics systems in astronomy \cite{HOL06}. Loss mechanisms of HCPCFs are coupling of the
 fundamental mode to interface modes due to diffuse scattering at glass/air interfaces and leakage
 of light to the exterior through the photonic crystal structure \cite{HOL06}. Another problem is
 the usually narrow transmission bandwidth \cite{COU06}.

When producing a fiber it is therefore very important to be able to optimize the fiber design to
 match desired properties and minimize undesired losses. Because of the complicated structure of
 PCFs this can only be done numerically. The oldest approach for computation of modes in fibers 
is the plane-wave expansion (PWE) method. Because of the complicated structure and great number of
 refractive index jumps in the fiber cross section this method is inefficient.
 For example discontinuous behaviour of the propagation modes at the glass/air interfaces can only
 be described with a huge number of basis functions taken into account for computation. This leads
 to very large computation times and poor convergence behaviour of the method \cite{HOL06}. Moreover,
 the modeling of an infinite exterior and therefore the simulation of radiation leakage is very
 difficult with the PWE method. 
Here we present the finite-element method (FEM) for the computation of leaky modes in photonic crystal fibers.

\section{Formulation of the propagation mode problem}
In this section we will derive the mathematical formulation of the propagation mode problem. The geometry of a fiber is invariant in one spacial dimension (along the fiber), here $z$-direction. A propagating mode is a solution to the time harmonic Maxwell's equations, which exhibits a harmonic dependency in $z$-direction:
\begin{eqnarray}
\MyField{E} & = & \MyField{E}_{\mathrm{pm}}(x, y)\exp \left(ik_{z}z\right)\nonumber\\
\MyField{H} & = & \MyField{H}_{\mathrm{pm}}(x, y)\exp \left(ik_{z}z\right)\label{eq:propAnsatz}. 
\end{eqnarray}
$\MyField{E}_{\mathrm{pm}}(x, y)$ and $\MyField{H}_{\mathrm{pm}}(x, y)$ are the electric and magnetic propagation modes and the parameter $k_{z}$ is called propagation constant. If the permittivity $\MyTensor{\epsilon}$ and permeability $\MyTensor{\mu}$ can be written as:
\begin{equation}
\MyTensor{\epsilon} =
\left[
\begin{array}{cc}
\MyTensor{\epsilon}_{\perp\,\perp} & 0\\
0 & \MyTensor{\epsilon}_{zz} 
\end{array}
\right]  
\quad \mbox{and} \quad 
\MyTensor{\mu} =
\left[
\begin{array}{cc}
\MyTensor{\mu}_{\perp\,\perp} & 0 \\
0 & \MyTensor{\mu}_{zz} \label{eq:permitPermea}
\end{array}
\right],
\end{equation}
we can split the propagation mode into a transversal and longitudinal component:
\begin{equation}
\MyField{H}_{\mathrm{pm}}(x, y) = 
\left[
\begin{array}{c}\MyField{H}_{\perp}(x, y) \\ \MySField{H}_{z}(x, y) \end{array}
\right].\label{eq:hprop}
\end{equation}
Inserting (\ref{eq:propAnsatz}) with (\ref{eq:permitPermea}) and (\ref{eq:hprop}) into Maxwell's equations yields:
\begin{equation}
\left[
\begin{array}{cc}
\MyMatrix{P} \nabla_{\perp} \MyTensor{\epsilon}_{zz}^{-1} \nabla_{\perp} \cdot \MyMatrix{P} 
-k_z^2 \MyMatrix{P} \MyTensor{\epsilon}_{\perp\,\perp}^{-1} \MyMatrix{P}\, &
-ik_z \MyMatrix{P} \MyTensor{\epsilon}_{\perp\,\perp}^{-1} \MyMatrix{P} \nabla_{\perp} \\
-ik_z\nabla_{\perp}\cdot \MyMatrix{P} \MyTensor{\epsilon}_{\perp\,\perp}^{-1} \MyMatrix{P} &
\nabla_{\perp}\cdot \MyMatrix{P} \MyTensor{\epsilon}_{\perp\,\perp}^{-1} \MyMatrix{P} \nabla_{\perp}
\end{array}
\right]
\left[
\begin{array}{c}\MyField{H}_{\perp} \\ \MySField{H}_{z} \end{array}
\right] =
\left[
\begin{array}{cc}
\omega^{2}\MyTensor{\mu}_{\perp\,\perp} & 0 \\
0 & \omega^{2}\MyTensor{\mu}_{zz}
\end{array}
\right]
\left[
\begin{array}{c}\MyField{H}_{\perp} \\ \MySField{H}_{z} \end{array}
\right],\label{eq:evp}
\end{equation}
with
\begin{equation}
\MyMatrix{P} =
\left[
\begin{array}{cc}
0 & -1 \\
1 & 0 
\end{array}
\right]
, \quad 
\nabla_{\perp} = 
\left[
\begin{array}{c} 
\partial_x \\
\partial_y
\end{array} 
\right]. 
\end{equation}
Now we define $\tilde{\MySField{H}}_{z}=k_{z}\MySField{H}_{z}$ and get:
\begin{equation}
\left[
\begin{array}{cc}
\MyMatrix{P} \nabla_{\perp} \MyTensor{\epsilon}_{zz}^{-1} \nabla_{\perp} \cdot \MyMatrix{P} 
-\omega^{2}\MyTensor{\mu}_{\perp\,\perp}\, &
-i \MyMatrix{P} \MyTensor{\epsilon}_{\perp\,\perp}^{-1} \MyMatrix{P} \nabla_{\perp} \\
0 &
\nabla_{\perp}\cdot \MyMatrix{P} \MyTensor{\epsilon}_{\perp\,\perp}^{-1} \MyMatrix{P} \nabla_{\perp}- \omega^{2}\MyTensor{\mu}_{zz}
\end{array}
\right]
\left[
\begin{array}{c}\MyField{H}_{\perp} \\ \tilde{\MySField{H}}_{z} \end{array}
\right] =k_z^2 
\left[
\begin{array}{cc}
\MyMatrix{P} \MyTensor{\epsilon}_{\perp\,\perp}^{-1} \MyMatrix{P}
 & 0 \\
i\nabla_{\perp}\cdot \MyMatrix{P} \MyTensor{\epsilon}_{\perp\,\perp}^{-1} \MyMatrix{P} &
0
\end{array}
\right]
\left[
\begin{array}{c}\MyField{H}_{\perp} \\ \tilde{\MySField{H}}_{z} \end{array}
\right]\quad x \in {\mathbb R}^{2}.  \label{eq:evp1}
\end{equation}
Eq. (\ref{eq:evp1}) is a quadratic eigenvalue problem for the propagation constant $k_{z}$ and propagation mode $\MyField{H}_{\mathrm{pm}}(x, y)$. We get a similar equation for $\MyField{E}_{\mathrm{pm}}(x, y)$ exchanging $\MyTensor{\epsilon}$ and $\MyTensor{\mu}$. For our numerical analysis we will not look at the propagation constant but define the effective refractive index $n_{\mathrm{eff}}$ which we will also refer to as eigenvalue:
\begin{equation}
\label{eq:neff}
n_{\mathrm{eff}} = \frac{k_z}{k_0} \qquad \mbox{with} \quad k_0 = \frac{2\pi}{\lambda_0},
\end{equation}
where $\lambda_{0}$ is the vacuum wavelength of light.
\section{Solution of the propagation mode problem with the finite-element method}
\begin{figure}[ht]
(a)\hspace{9cm}(b)\\
\includegraphics[width=7cm]{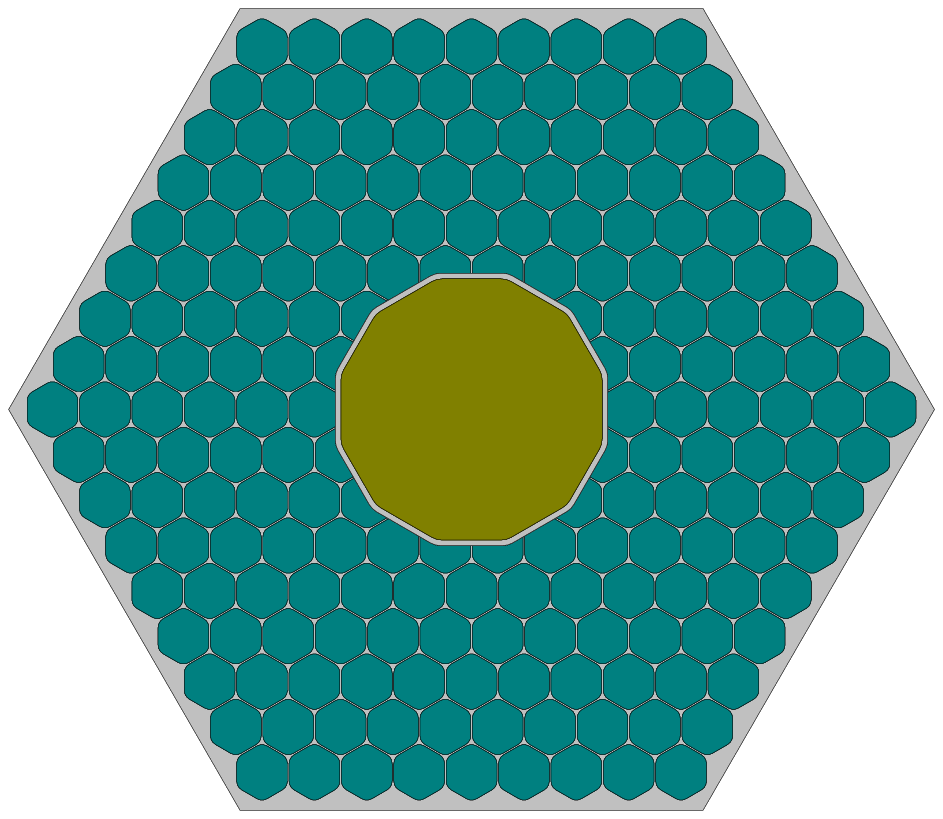}\hfill
\includegraphics[width=7cm]{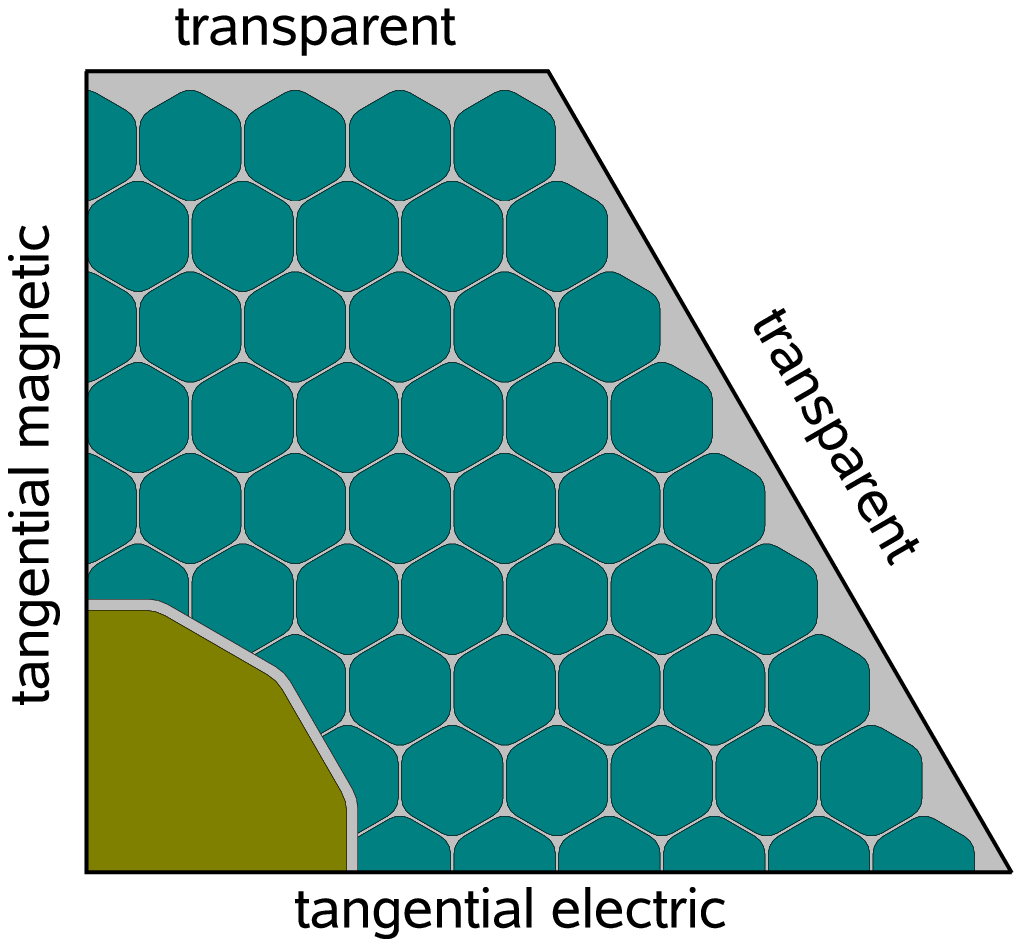}\\
(c)\hspace{9cm}(d)\\
\includegraphics[width=7cm]{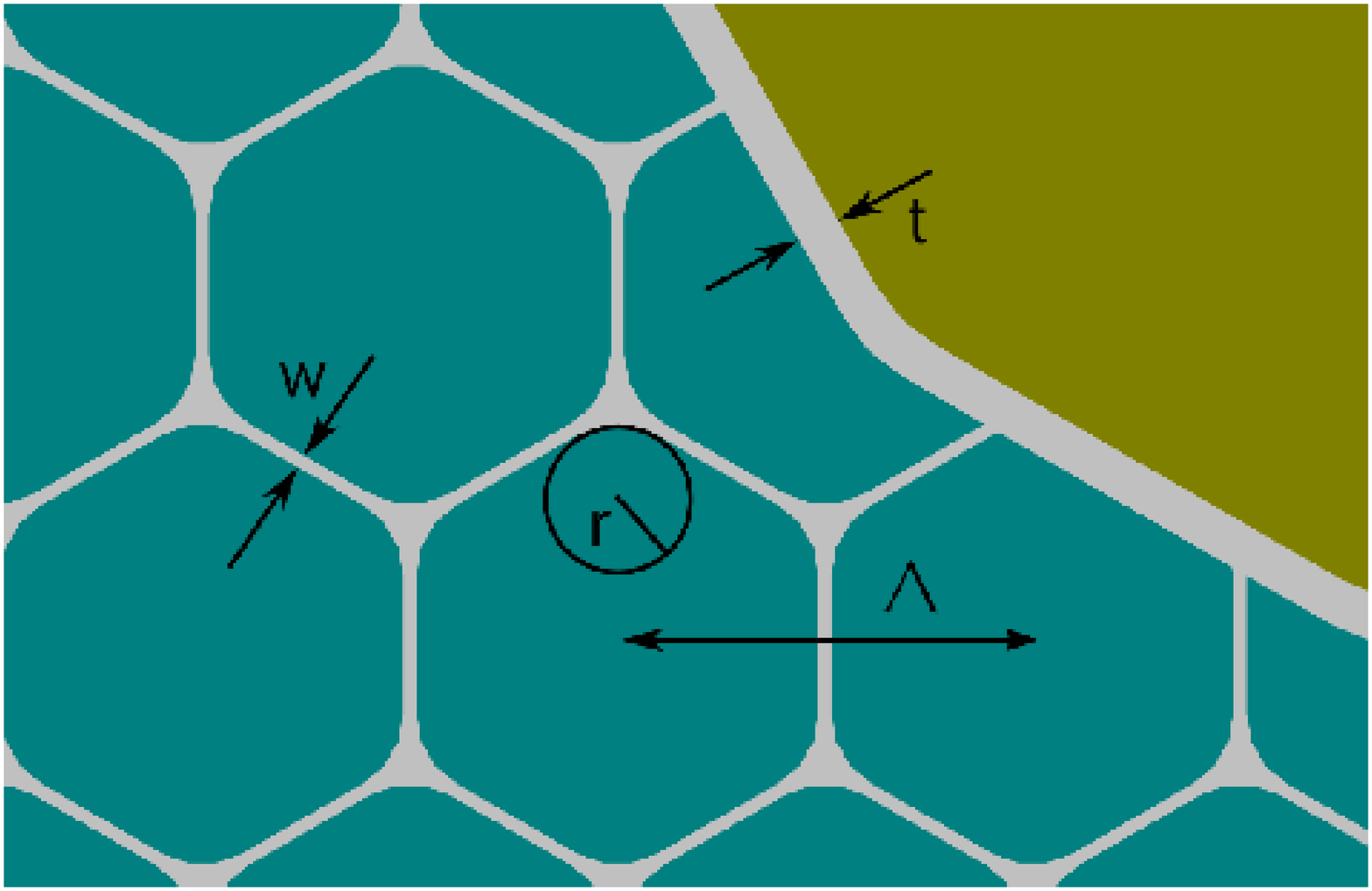}\hfill
\includegraphics[width=7cm]{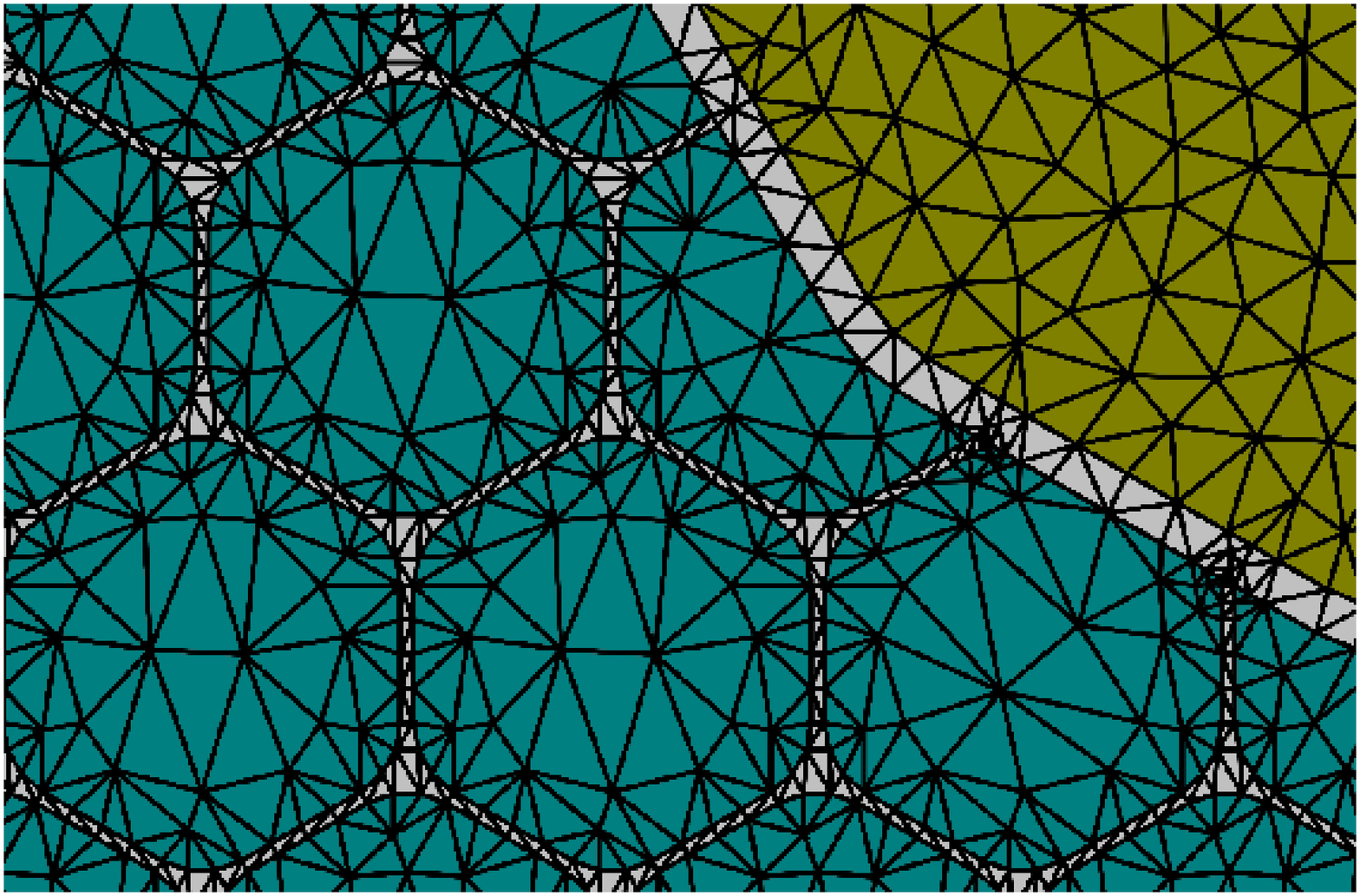}
\caption{\label{fig:hcpcfTriang}(a) Geometry of HCPCF used for mode computation; (b) boundary conditions for mode computation with a quarter of the fiber; (c) geometrical parameters describing HCPCF. Pitch $\Lambda$, hole edge radius $r$, strut thickness $w$, core surround thickness $t$; (d) detail from a triangulation of HCPCF. Due to the flexibility of triangulations all geometrical features of the HCPCF are resolved.}
\end{figure}
\begin{figure}[ht]
(a)\hspace{5.5cm}(b)\hspace{5.5cm}(c)\\
\includegraphics[width=5.2cm,height=5.2cm]{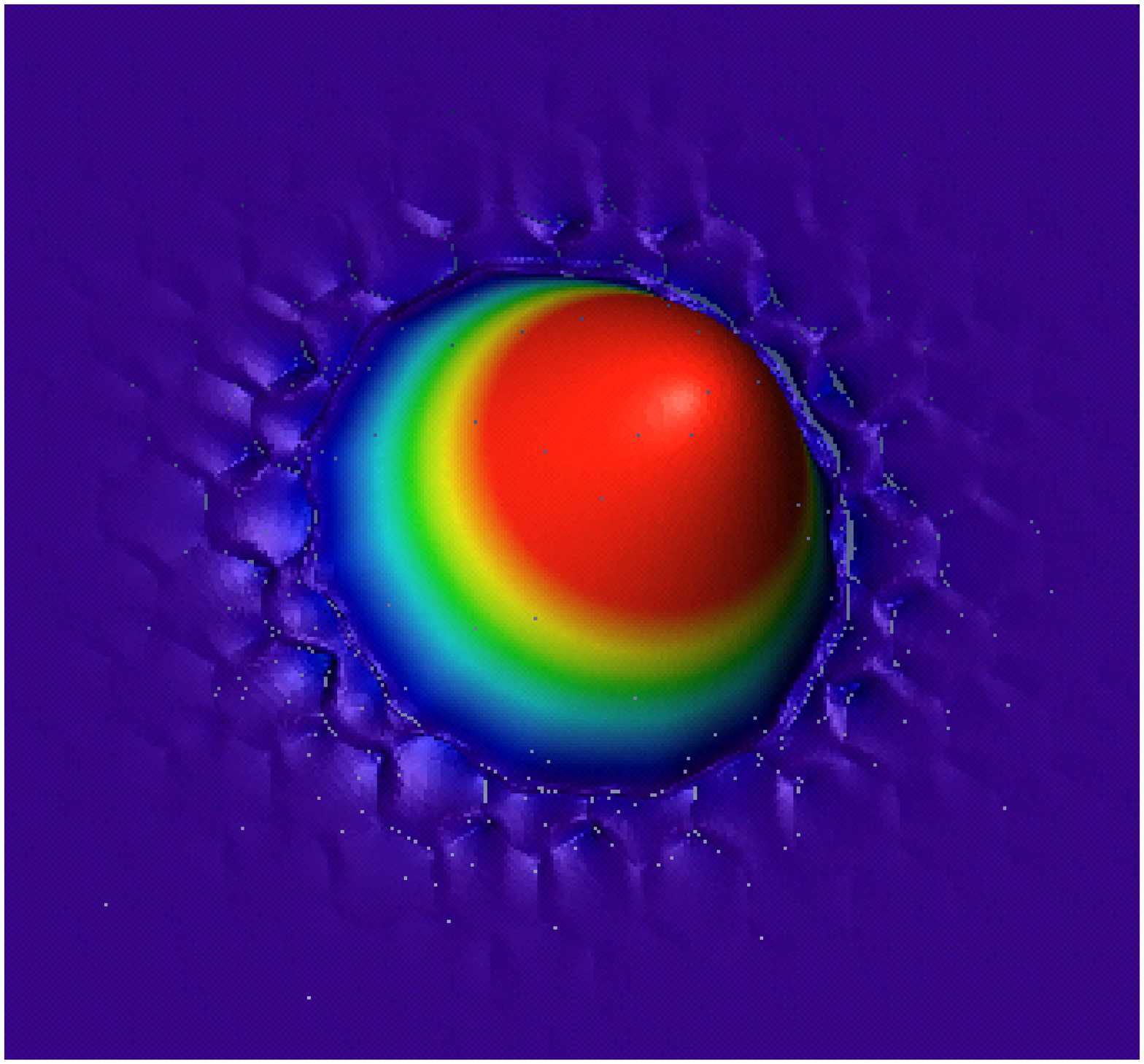}\hfill
\includegraphics[width=5.2cm,height=5.2cm]{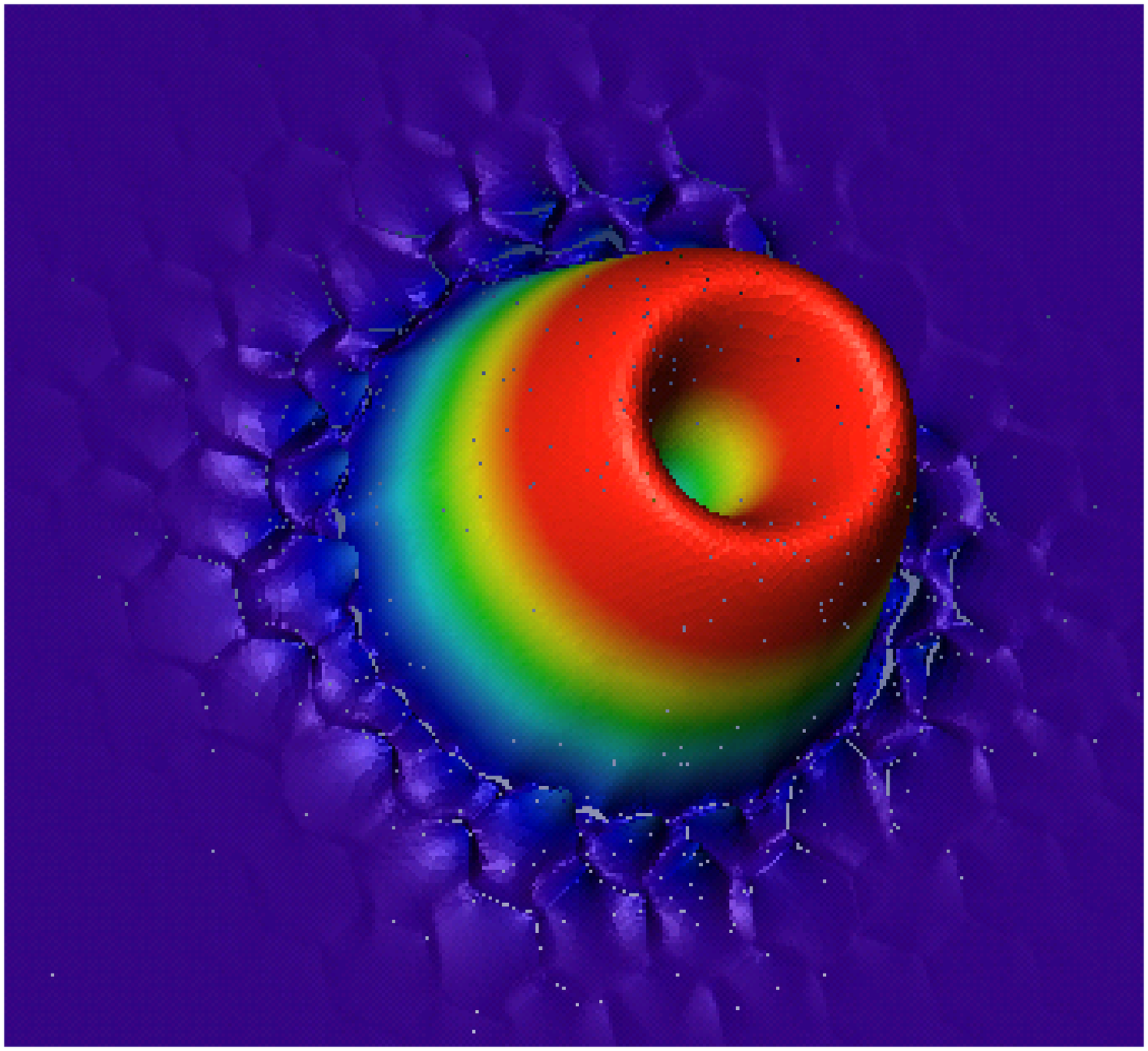}\hfill
\includegraphics[width=5.2cm,height=5.2cm]{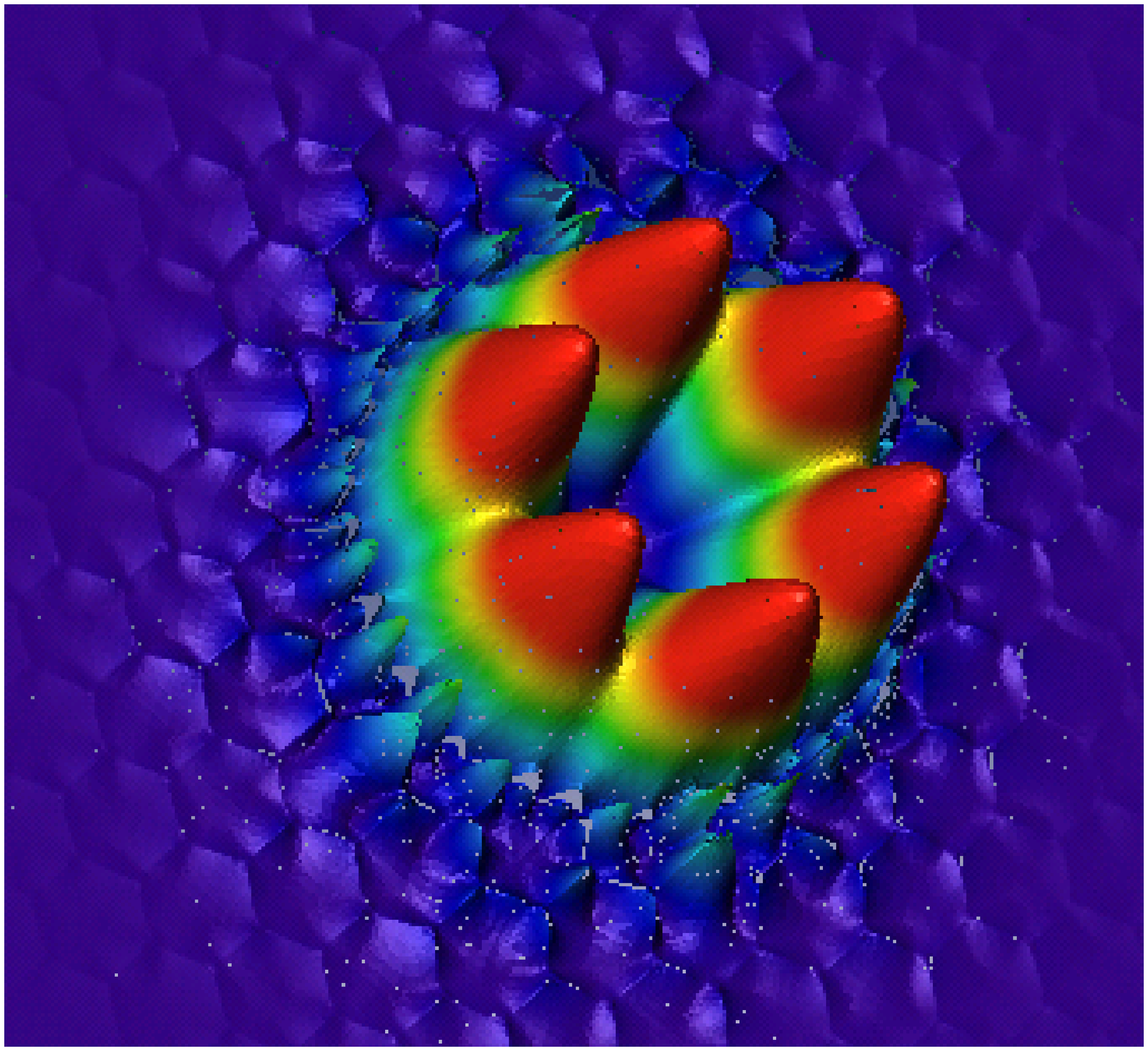}\hfill
\caption{\label{fig:coreModes}First, second and fourth fundamental core modes of HCPCF illustrated in Fig. \ref{fig:hcpcfTriang}(a) - Parameters: $\Lambda=1550\,$nm, $r=300\,$nm, $w=50\,$nm, $t=170\,$nm.}
\end{figure}
In order to compute propagation modes we have to solve the eigenvalue problem (\ref{eq:evp1}) numerically. Note that this problem is formulated on ${\mathbb R}^{2}$. We have an eigenvalue problem on an unbounded domain.
Usually, propagation modes are computed on an artificially bounded domain $\Omega$ applying either periodic or so called Dirichlet boundary conditions, i.e. $\MyField{E}_{\mathrm{pm}}(x, y)=0$ for $(x,y)\in\partial\Omega$ (propagation mode vanishes on the boundary $\partial\Omega$ of the computational domain). Using this simplification it is not possible to model leaking of light from the core of the fiber to the exterior (leaky modes). Here we want to take this effect into account. Since our computational domain still has to be of
 finite size, we apply so-called transparent boundary conditions to $\partial\Omega$. 
We realize these boundary conditions with the perfectly matched layer (PML) method \cite{BerPML}. 
The propagation constant $k_{z}$ becomes complex and the corresponding mode is damped according
 to ${\mathrm{exp}}({-\Im(k_{z})z})$ while propagating along the fiber. This dampening is due to radiation leakage from the fiber to the exterior.

In contrast to the PWE method, whose ansatz functions are spread over the whole computational domain (plane waves) the FEM method uses localized ansatz functions. To construct these ansatz functions, the computational domain has to be discretized. This means the geometry is subdivided into $N$ patches, e.g. triangles in two dimensions, tetrahedra in three dimensions. Fig. \ref{fig:hcpcfTriang}(d) shows such a triangulation of a photonic crystal fiber (Fig. \ref{fig:hcpcfTriang}(c)). On the patches usually polynomial ansatz functions are defined. Since we are solving Maxwell's equations, our solution is a vectorial function and therefore we use vectorial ansatz functions $\MyField{\nu}_{i}(x,y)$. The numerical solution for the electric field $\tilde{\MyField{E}}$ is a superposition of these localized ansatz functions
\begin{equation}
  \label{eq:fem_super}
  \tilde{\MyField{E}}(x,y)=\sum_{i=1}^{N}h_{i} \MyField{\nu}_{i}(x,y)
\end{equation}
The FEM computation determines the unknown coefficients $h_{i}$. The FEM method has several advantages:
\begin{itemize}
\item Maxwell's equations are solved rigorously without approximations.
\item The flexibility of triangulations allows the computation of virtually arbitrary structures without simplifications or approximations, as illustrated in Fig. \ref{fig:hcpcfTriang}(d).
\item Choosing appropriate ansatz functions $\MyField{\nu}_{i}(x,y)$ for the solution of Maxwell's equations, physical properties of the electric field like discontinuities or singularities can be modeled very accurately and don't give rise to numerical problems. Such discontinuities often appear at glass/air interfaces of photonic crystal fibers, see Fig. \ref{fig:coreModes}(c).
\item Adaptive mesh-refinement strategies lead to very accurate results and small computational times which is a crucial point for optimization of fiber design.
\item The FEM approach converges with a fixed convergence rate towards the exact solution of Maxwell-type problems for decreasing mesh width (i.e. increasing number $N$ of sub-patches) of the triangulation. Therefore, it is easy to check if numerical results can be trusted \cite{MON03}.
\end{itemize}
In the following sections we apply the FEM method to the computation of leaky propagation modes in HCPCFs. The only simplification we make is to extend the fiber cladding to infinity and thereby neglect its finite size. This is justified if the cladding of the fiber is much larger than the microstructured core, and no light entering the cladding is reflected back into the core, which is usually the case. 

Throughout this paper we use the FEM package JCMsuite developed at the Zuse Institute Berlin for the numerical solution of Maxwell's equations. It has been successfully applied to a wide range of electromagnetic field computations including waveguide structures \cite{Burger2005a}, DUV phase masks \cite{Burger2005bacus}, and other nano-structured materials \cite{Enkrich2005a,Kalkbrenner2005a}. It provides higher-order edge elements, multi-grid methods, a-posteriori error control, adaptive mesh refinement, and adaptive transparent boundary conditions. Further numerical details about the computation of leaky modes with JCMsuite can be found in \cite{Zschiedrich2005a}.

\section{Numerical aspects of propagation mode computation}
\begin{figure}[ht]
\psfrag{p1}{p=1}
\psfrag{p2}{p=2}
\psfrag{p3}{p=3}
\psfrag{p4}{p=4}
\psfrag{p5}{p=5}
\psfrag{p6}{p=6}
\psfrag{p7}{p=7}
\psfrag{unk}{unknowns}
\psfrag{ada2}{p=2, adaptive}
\psfrag{ada4}{p=4, adaptive}
\psfrag{ada5}{p=5, adaptive}
\psfrag{no2}{p=2, uniform}
\psfrag{no4}{p=4, uniform}
\psfrag{no5}{p=5, uniform}
\psfrag{deltare}{$\Delta[\Re(n_{\mathrm{eff}})]$}
\psfrag{deltaim}{$\Delta[\Im(n_{\mathrm{eff}})]$}
(a)\hspace{9cm}(b)\\
\includegraphics[width=8cm]{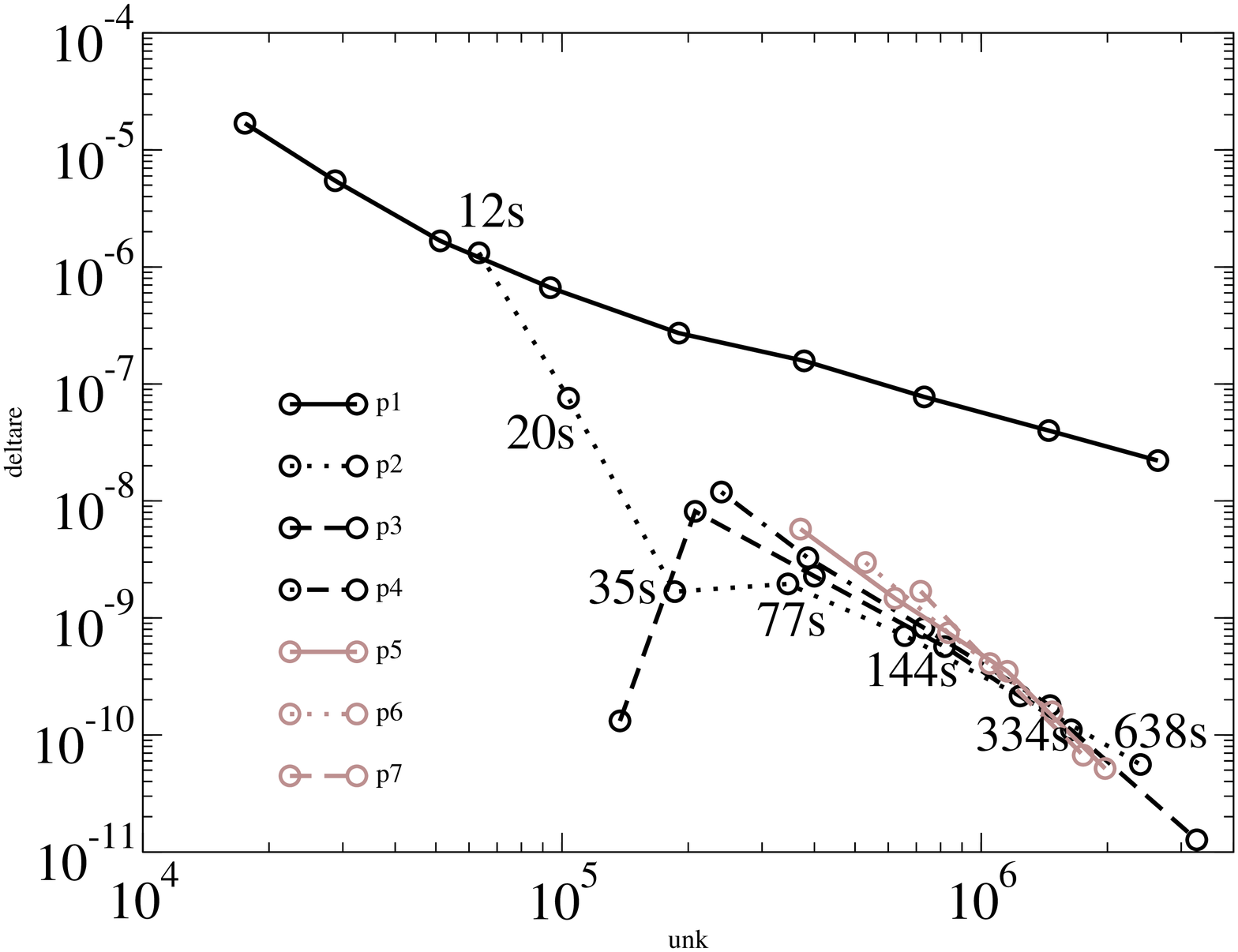}\hfill
\includegraphics[width=8cm]{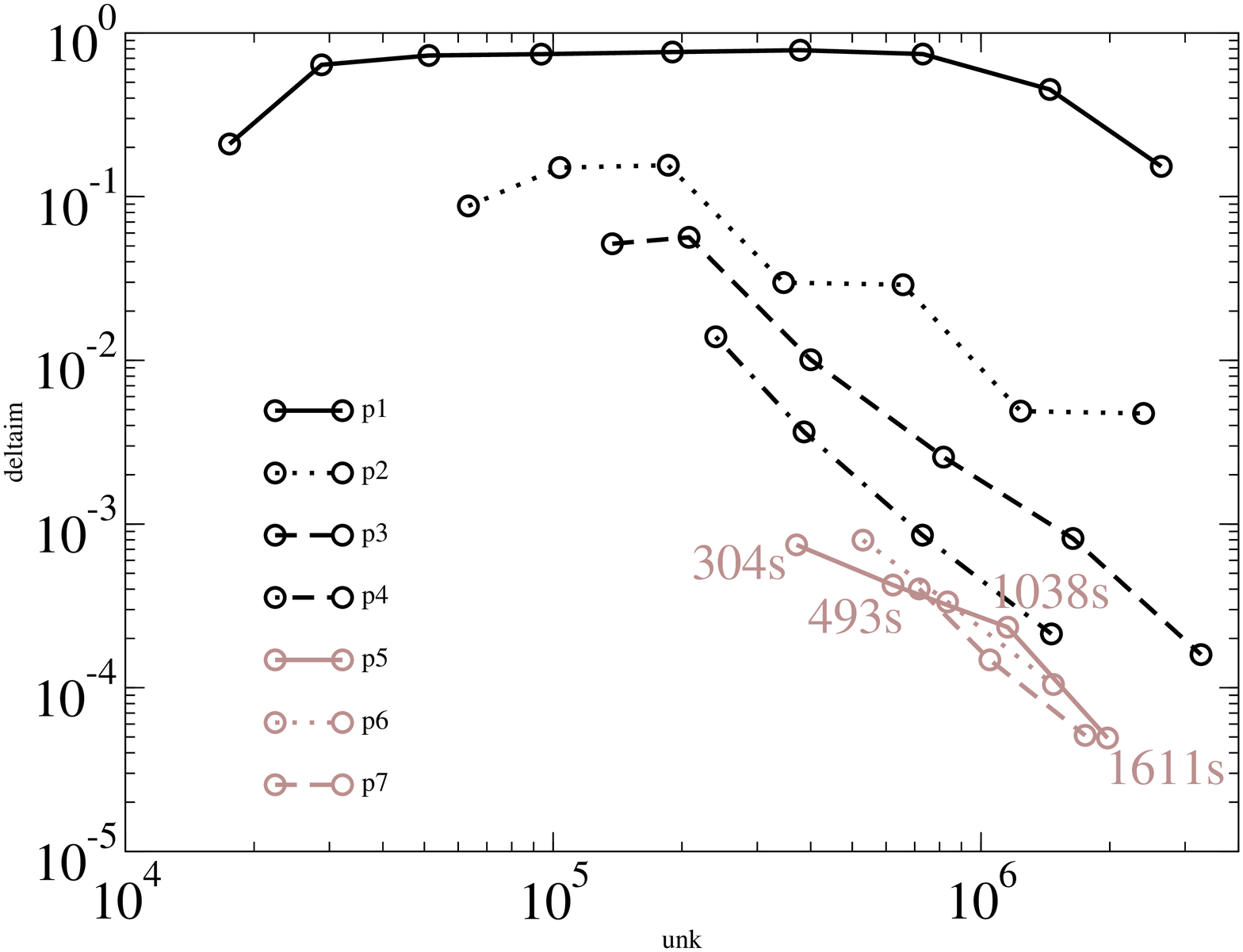}\\
(c)\hspace{9cm}(d)\\
\includegraphics[width=8cm]{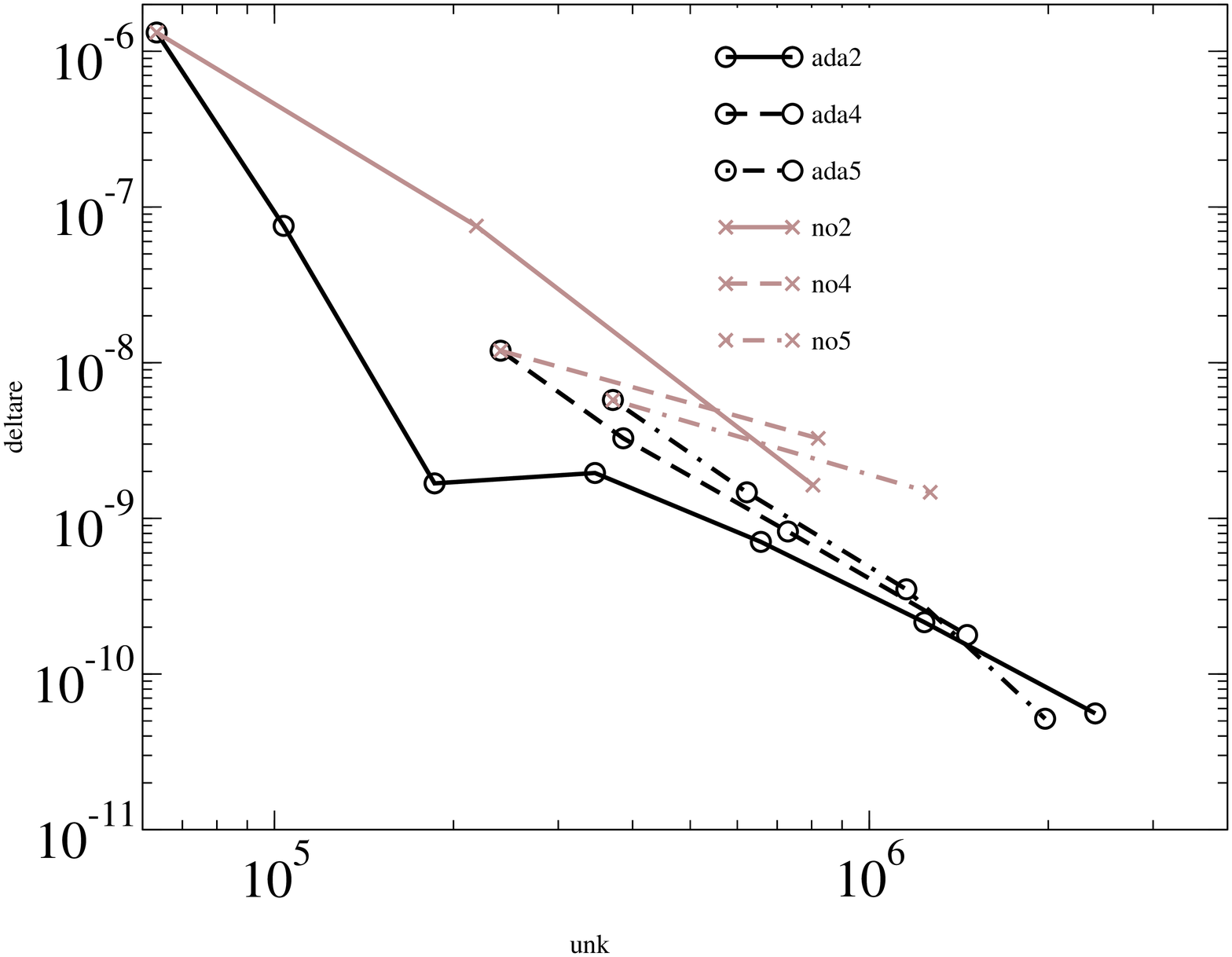}\hfill
\includegraphics[width=8cm]{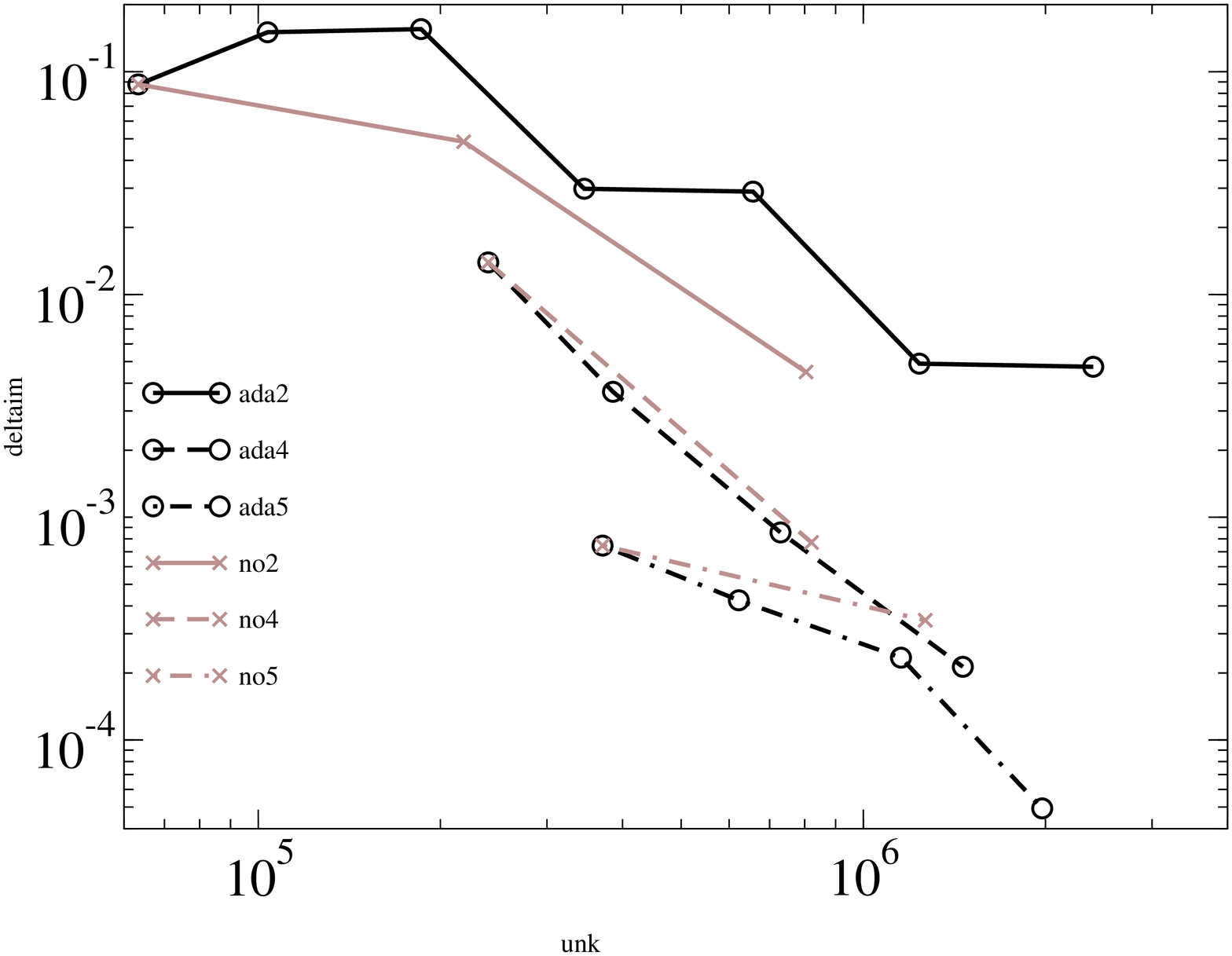}\\
\caption{\label{fig:convAda}Relative error of first eigenmode in dependence on number of unknowns of FEM computation for different FEM degrees\ $p$. (a) and (b): adaptive refinement; (c) and (d) comparison of adaptive and uniform refinement. Parameters: $\Lambda=1550\,$nm, $r=300\,$nm, $w=50\,$nm, $t=170\,$nm.}
\end{figure}

Fig. \ref{fig:coreModes} shows the first, second, and fourth core modes with lowest energy of the HCPCF illustrated in Fig. \ref{fig:hcpcfTriang}(a). All of these core modes appear several times in the spectrum because of the $\mathrm{C}_{6\mathrm{V}}$ symmetry of the layout. To decrease the problem size, we only use one quarter of the fiber as computational domain. Of course we should obtain the same eigenvalues taking the full, half or quarter fiber as computational domain, as demonstrated in Table \ref{table:eigenvaluesSymmetry}. The imaginary part of the eigenvalue is much smaller than the real part. The differences in $\Im(n_{\mathrm{eff}})$ are therefore due to the numerical uncertainty at the chosen refinement level. $\Im(n_{\mathrm{eff}})$  will of course converge with increasing number of ansatz functions, see Fig. \ref{fig:convAda}.
Figure \ref{fig:hcpcfTriang}(b) shows which boundary conditions we apply to the artificial inner boundaries for the case of a quarter fiber. These boundary conditions follow from symmetry of the core modes.
\begin{table}
\centering
\begin{tabular}{|l|c|c|c|c|}
\hline
&unknowns & 1st eigenvalue & 2nd eigenvalue & 3rd eigenvalue \\
\hline
full fiber& 861289    & $0.998265726 + 9.508\cdot 10^{-12}i$ &$0.99212759 + 1.513\cdot 10^{-10}i$   &$0.9908968+ 2.089\cdot 10^{-10}i$ \\
\hline
half fiber & 438297   & $0.998265719 + 9.357\cdot 10^{-12}i$ &$0.99212742 + 1.489\cdot 10^{-10}i$   &$0.9908958+ 2.086\cdot 10^{-10}i$\\
\hline
quarter fiber & 218504 & $0.998265724 + 9.372\cdot 10^{-12}i$ &$0.99212652 + 1.504\cdot 10^{-10}i$   &$0.9909169+ 2.302\cdot 10^{-10}i$ \\
\hline
\end{tabular}
\caption{\label{table:eigenvaluesSymmetry}First, second and third eigenvalue computed with full, half and quarter fiber as computational domain.}
\end{table}
Note that the imaginary and real parts of the eigenvalues differ by 11 orders of magnitude. Computation of leaky modes is therefore a multi-scale problem which makes it numerically very difficult. When analyzing radiation losses of a fiber, the imaginary part of the effective refractive index $n_{\mathrm{eff}}$ is the quantity of interest. 
In the following we analyze how accurately we can compute $n_{\mathrm{eff}}$, i.e. we look at the convergence of the FEM computation. Therefore we compute $n_{\mathrm{eff}}$ with an increasing number of unknowns. We can either refine the triangulation or increase the polynomial degree $p$ of the ansatz functions on each patch to obtain a more accurate numerical result. The software package JCMsuite offers finite elements with a maximum degree of $9$. Furthermore, the refinement of the grid can be performed uniformly or adaptively. In a uniform refinement step each triangle is subdivided into 4 smaller ones. In an adaptive refinement step, an a-posteriori error estimator automatically chooses a certain number of triangles where the error of the FEM solution is large. Only these triangles are refined. This method leads to accurate results with smaller number of unknowns which also implies shorter computational times and smaller memory requirements.
Fig. \ref{fig:convAda} shows the relative error of the real and imaginary parts of $n_{\mathrm{eff}}$ in dependence on the number of unknowns of the FEM computation for different FEM degrees\ $p$
. The relative error decreases with an increasing number of unknowns. Let us first look at the convergence with adaptive refinement \ref{fig:convAda}(a), (b). For the real part we achieve the fastest convergence for a finite element degree of $2$. Note that already for a finite element degree of $1$ and only $20,000$ unknowns we have a relative error smaller than $10^{-5}$. For $10^{6}$ unknowns we have a relative error smaller than $10^{-9}$ for FEM degrees greater than $1$. The imaginary part of the effective refractive index is much harder to compute accurately. For the same number of unknowns its relative error is much larger than the relative error of the real part. For a FEM degree of $1$ the imaginary part does not converge at all up to $3\cdot10^{6}$ unknowns. However the higher the FEM degree the faster the imaginary part converges. For the computation of leaky modes in HCPCFs it is therefore important to employ higher-order finte elements. 
Fig. \ref{fig:convAda}(c), (d) shows a comparison of the convergence of the real and imaginary parts between adaptive and uniform refinement. While adaptive refinement leads to a faster convergence of the real part for all FEM degrees the imaginary part converges almost equally fast for an FEM degree of 4 and even slower for an FEM degree of 2. This happens because the imaginary part of the eigenvalue is much smaller than the real part. The error estimator which chooses triangles in the case of adaptive refinement primarily reduces the error of the real part first, since its contribution to the relative error of the complex eigenvalue is much larger than the imaginary part. Computing leaky modes a uniform refinement strategy is therefore more suitable if one is interested in the imaginary part and radiation losses. 

For further numerical analysis of radiation losses of the HCPCF we choose a finite-element degree of $6$ and no refinement step. With these settings the numerical problem has approximately $500,000$ unknowns for a triangulation according to Fig. \ref{fig:hcpcfTriang}(d). The relative error of the real and imaginary parts is $\approx 10^{-8}$ and $\approx10^{-3}$ respectively, according to the convergence curves. Such a computation requires $2$GB RAM and about 10 minutes computation time on a standard desktop PC. It is fast enough for multidimensional optimization of the fiber design.
\section{Optimization of HCPCF design}
\begin{figure}[ht]
\psfrag{neffimag}{$\Im(n_{\mathrm{eff}})$}
\psfrag{lambda}{$\Lambda$}
\psfrag{wlabel}{$w$}
\psfrag{core}{$t$}
\psfrag{nrows}{cladding rings}
\psfrag{hole}{$r$}
(a)\hspace{5.5cm}(b)\hspace{5.5cm}(c)\\
\includegraphics[width=5.3cm]{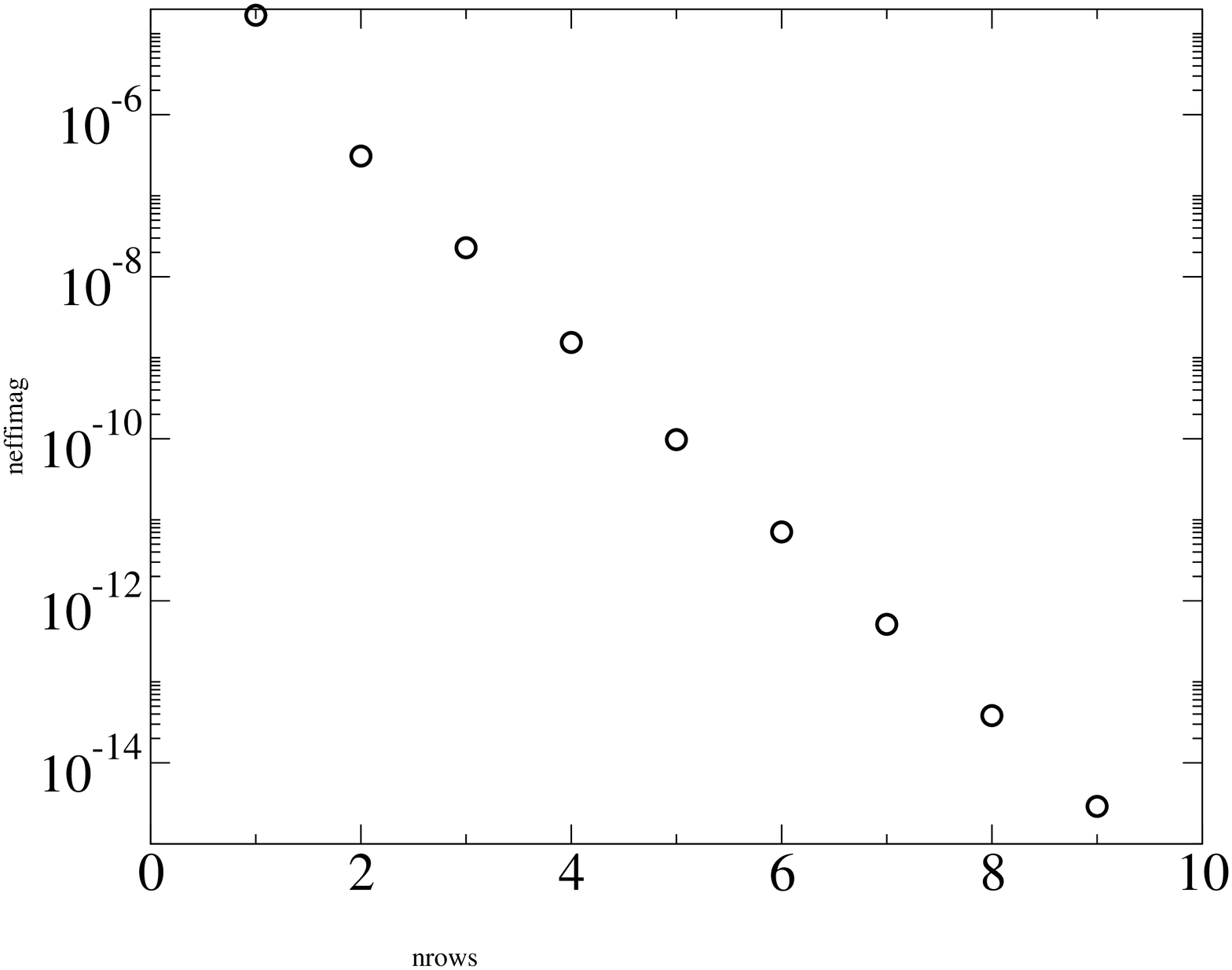}\hfill
\includegraphics[width=5.3cm]{fig/pitchScan2_.eps}\hfill
\includegraphics[width=5.3cm]{fig/coreSurroundScan_.eps}\hfill
\\
(d)\hspace{5.5cm}(e)\\
\includegraphics[width=5.3cm]{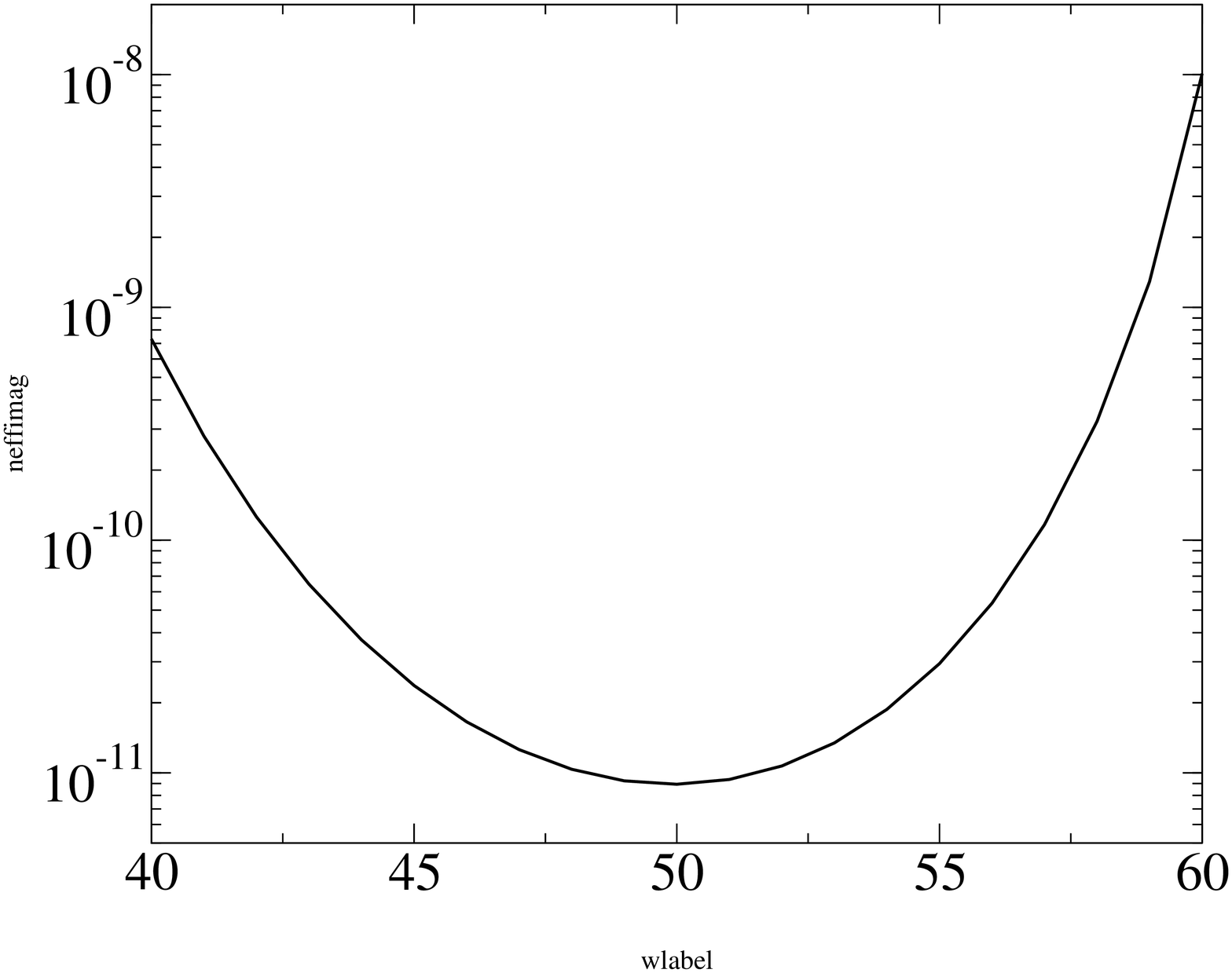}\hspace{5mm}
\includegraphics[width=5.3cm]{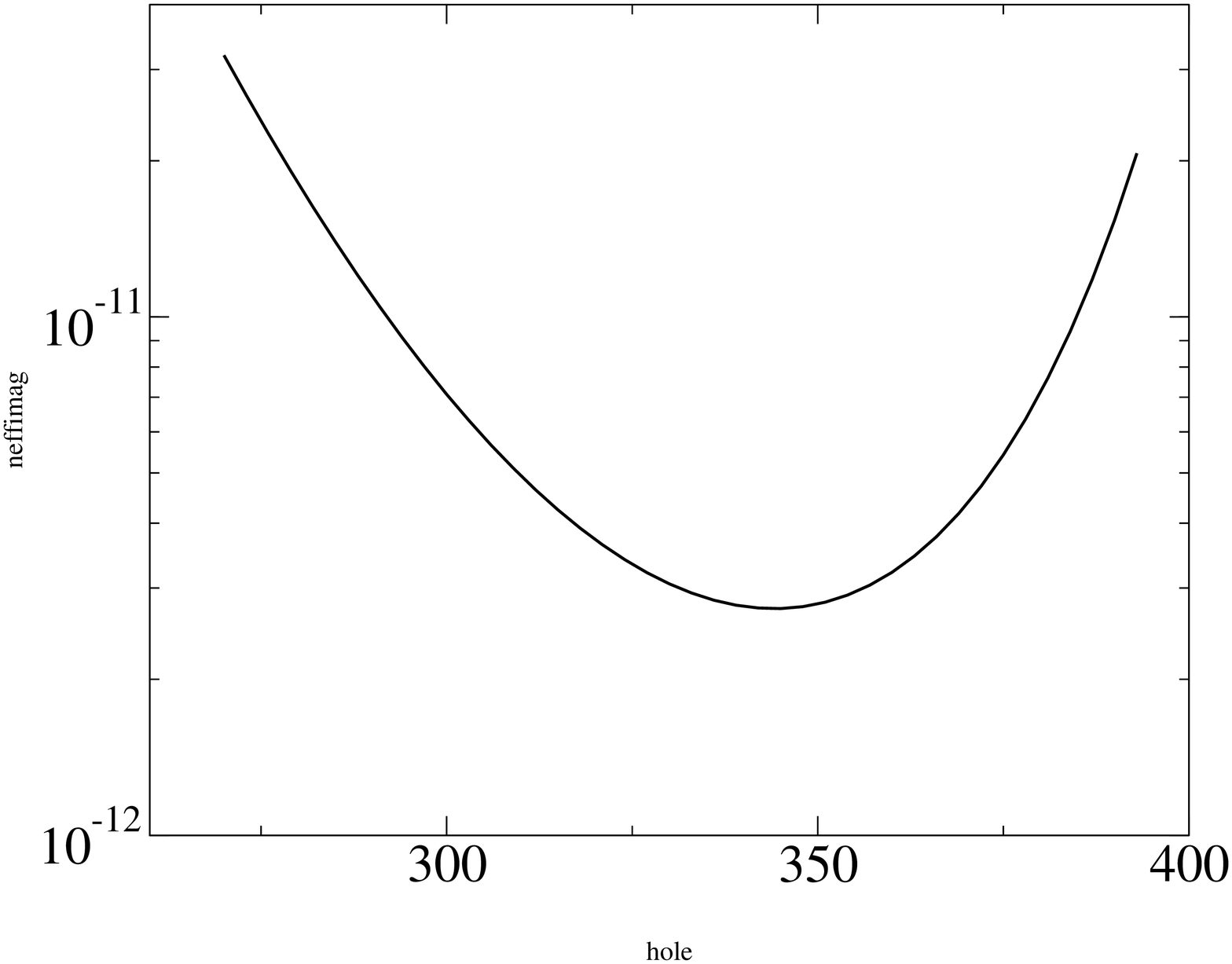}
\caption{\label{fig:geoScan}Imaginary part of effective refractive index $\Im(n_{\mathrm{eff}})$ in dependence on: (a) number of cladding rings, (b) pitch $\Lambda$, (c) core surround thickness $t$, (d) strut thickness $w$, (e) hole edge radius $r$. Parameters: $\Lambda=1550\,$nm, $r=300\,$nm, $w=50\,$nm, $t=170\,$nm, 6 cladding rings, wavelength $\lambda=589\,$nm.}
\end{figure}

In this section we want to optimize the fiber design shown in figure \ref{fig:hcpcfTriang}(a) in order to reduce radiation losses, i.e. we want to minimize $\Im(n_{\mathrm{eff}})$. The basic fiber layout is a 19-cell core with rings of hexagonal cladding cells. Since these cladding rings prevent leakage of radiation to the exterior we expect that an increasing number of cladding rings reduces radiation leakage and therefore $\Im(n_{\mathrm{eff}})$. This is confirmed by our numerical simulations shown in Fig. \ref{fig:geoScan}. The radiation leakage decreases exponentially with the number of cladding rings and thereby the thickness of the photonic crystal structure. This behavior agrees with the exponential dampening of light propagating through a photonic crystal structure with frequency in the photonic band gap.

For our further analysis we fix the number of cladding rings to 6. The free geometrical parameters are the pitch $\Lambda$, hole edge radius $r$, strut thickness $w$, and core surround thickness $t$, see Fig. \ref{fig:hcpcfTriang}(c).
Fig. \ref{fig:geoScan} shows the imaginary part of the effective refractive index in dependence on these parameters. For each scan all but one parameter were fixed. For the strut thickness $w$ and the hole edge radius $r$ we find well-defined optimal values which minimize $\Im(n_{\mathrm{eff}})$. For pitch $\Lambda$ and core surround thickness $t$ a large number of local minima and maxima can be seen. Now we want to optimize the fiber design using multidimensional optimization with the Nelder-Mead simplex method\cite{HOL06}. To reduce the number of optimization parameters we fix the hole edge radius to the determined minimum at $r=354\,$nm since its variation has the smallest effect on $\Im(n_{\mathrm{eff}})$. For optimization we have to choose starting values for $\Lambda$, $t$ and $w$. Since the simplex method searches for local minima we have to decide in which local minimum of $\Lambda$ and $t$ we want to search. We choose $t=152\,$nm since here $\Im(n_{\mathrm{eff}})$ has a global minimum and $\Lambda=1550\,$nm since the bandwidth of this minimum is much larger than for the global minimum at $\Lambda=1700\,$nm.

Optimization yields a minimum value of 
$\Im(n_{\mathrm{eff}})=5\cdot 10^{-15}\frac{1}{\mathrm{m}}$ for the imaginary part of the effective refractive index. The corresponding geometrical parameters are $\Lambda=1597\,$nm, $w=38\,$nm , $t=151\,$nm.

\section{Conclusion}
We have demonstrated that the finite-element method is very well suited for the analysis of light propagation in hollow-core photonic crystal fibers. Here we considered a 19-cell HCPCF with 6 cladding rings. A convergence analysis allowed us to quantify the relative error of real and imaginary parts of the propagation constant of leaky eigenmodes. The real part could be computed in $20\,$s on a standard desktop PC down to a relative error of $10^{-7}$. The imaginary part, being $10$ orders of magnitude smaller, could be determined in about $10\,$min with a relative error smaller than $10^{-3}$. Thanks to the short computational time, we were able to optimize the fiber design regarding pitch, strut thickness, core surround thickness, and hole edge radius of the cladding cells and therewith minimize radiation losses.
\section{Acknowledgement}
We acknowledge support by DFG within SP 1113. Furthermore we thank P.J. Roberts for fruitful discussions.

\bibliography{myBib}
\bibliographystyle{spiebib}

\end{document}